\documentclass[%
 reprint,
 amsmath,amssymb,
 aps,
superscriptaddress
]{revtex4-2}

\usepackage{graphicx}
\usepackage{dcolumn}
\usepackage{bm}
\usepackage{physics}
\usepackage{siunitx}
\usepackage{upgreek}
\usepackage{hyperref}
\usepackage[capitalise]{cleveref}
\usepackage{mathtools}
\usepackage{color}
\usepackage[markup=nocolor,deletedmarkup=xout]{changes}
\newcommand{\mueV}{\upmu eV}
\newcommand{\meV}{\text{meV}}
\renewcommand{\i}{\text{i}}
\newcommand{\Chi}{\raisebox{1.7pt}{$\chi$}}
\newcommand{\kb}{k_\text{B}}

\begin{document}

\preprint{APS/123-QED}

\title{Ultrafast electric control of cavity mediated single-photon and \\ photon-pair generation with semiconductor quantum dots}

\author{David Bauch}
\affiliation{Department of Physics and Center for Optoelectronics and Photonics Paderborn (CeOPP), Paderborn University, Warburger Strasse 100, 33098 Paderborn, Germany}

\author{Dirk Heinze}
\affiliation{Department of Physics and Center for Optoelectronics and Photonics Paderborn (CeOPP), Paderborn University, Warburger Strasse 100, 33098 Paderborn, Germany}

\author{Jens F\"orstner}
\affiliation{Electrical Engineering Department and Center for Optoelectronics and Photonics Paderborn (CeOPP), Paderborn University, Warburger Str. 100, 33098 Paderborn, Germany}

\author{Klaus D. J\"ons}
\affiliation{Department of Physics and Center for Optoelectronics and Photonics Paderborn (CeOPP), Paderborn University, Warburger Strasse 100, 33098 Paderborn, Germany}
\affiliation{Institute for Photonic Quantum Systems, Paderborn University, 33098 Paderborn, Germany}

\author{Stefan Schumacher}
\affiliation{Department of Physics and Center for Optoelectronics and Photonics Paderborn (CeOPP), Paderborn University, Warburger Strasse 100, 33098 Paderborn, Germany}

\affiliation{Wyant College of Optical Sciences, University of Arizona, Tucson, Arizona 85721, USA}

\date{\today}
             
\begin{abstract}
Employing the ultrafast control of electronic states of a semiconductor quantum dot in a cavity, we introduce a novel approach to achieve on-demand emission of single photons with almost perfect indistinguishability and photon pairs with near ideal entanglement. Our scheme is based on optical excitation off-resonant to a cavity mode followed by ultrafast control of the electronic states using the time-dependent quantum-confined Stark effect, which then allows for cavity-resonant emission. Our theoretical analysis takes into account cavity-loss mechanisms, the Stark effect, and phonon-induced dephasing allowing realistic predictions for finite temperatures. 
\end{abstract}

\maketitle

\section{\label{sec:introduction}Introduction}
Semiconductor quantum dots (QDs) are discussed as leading candidates for ideal on-demand generation of single photons and entangled photon pairs, with reportedly high indistinguishabilities, emission efficiencies, and purities \cite{Ding2016OnDemandSinglePhotons,Hanschke2018SinglePhoton,schweickert2018demand,Chen2018EntangledPhotonsEfficientExtractionAntenna,Huber2008StrainEntangledOnDemand,Liu2019EntangledEhotonBrightnessIndistinguishability,Wang2019OnDemandEntangledPhotonsFidelityEfficiencyIndistinguishability,huber2017highlySPTPSourceIndistConc,muller2014demand}. In most cases, losses and decoherence are reduced by using optical cavities to enhance and accelerate the photon emission. 
However, for both single-photon as well as degenerate twin-photon emission, efficient resonant excitation and resonant cavity-enhanced emission appear to be mutually exclusive. 
For example, resonant excitation of the quantum dot can reduce the effective brightness of the source as only photons in the cross polarized channel are collected \cite{NearOptimaSinglePhoton2016}, while the excitation pulse may also directly generate cavity photons, undermining the potential quality of the emitted photons. 
One possible way to evade this problem is to substitute the resonant excitation process using dichromatic pulses that excite an exciton that then resonantly emits a photon into a cavity mode \cite{DichromaticExcitation2019,DichromaticExcitation2021}. 
Another way to circumvent the resonant excitation is to indirectly excite the exciton via a phonon side band. However, this requires sufficiently high pulse areas to reliably prepare the exciton compared to a single, resonant $\pi$-pulse \cite{glassl2013proposed,quilter2015phonon,ardelt2014dissipative}. 
The emission of entangled photon pairs from a quantum dot biexciton is well understood and can be optimized with photonic structures to overcome the limited indistinguishability of the cascaded photons \cite{ESchoell2020,hafenbrak2007triggeredEntangledPhotons,huber2017highlySPTPSourceIndistConc}. 
Here, two-photon excitation \cite{StuflerTPE2006,ardelt2014dissipative,jayakumar2013deterministic} is typically used, exciting the biexciton state from the initial ground state without populating the exciton states. 
For a non-zero biexciton binding energy this excitation process is naturally off-resonant with the electronic single-photon transitions. 
This is, however, not true in the case of degenerate twin-photon emission at half the biexciton energy \cite{Ota2011TwoPhotonEmission}, where both the direct degenerate two-photon excitation of the biexciton and the two-photon emission would be resonant with the cavity mode. 
Photons emitted by this process are highly entangled and indistinguishable \cite{Schumacher12,heinze2017polarization,cygorek2018comparison}, but again the simultaneous resonant excitation and emission needs to be avoided. 
In the light of recent experimental and technological developments, which demonstrated fast electrical control of electronic resonances \cite{WarburtonLodahl2020} and coherent control of excitonic states \cite{Widhalm2018}, we propose and theoretically explore an optoelectronic scheme to excite the quantum dot exciton and biexciton, from which photons can then efficiently and resonantly be emitted into a cavity mode. 

\begin{figure}
\includegraphics[width=\columnwidth]{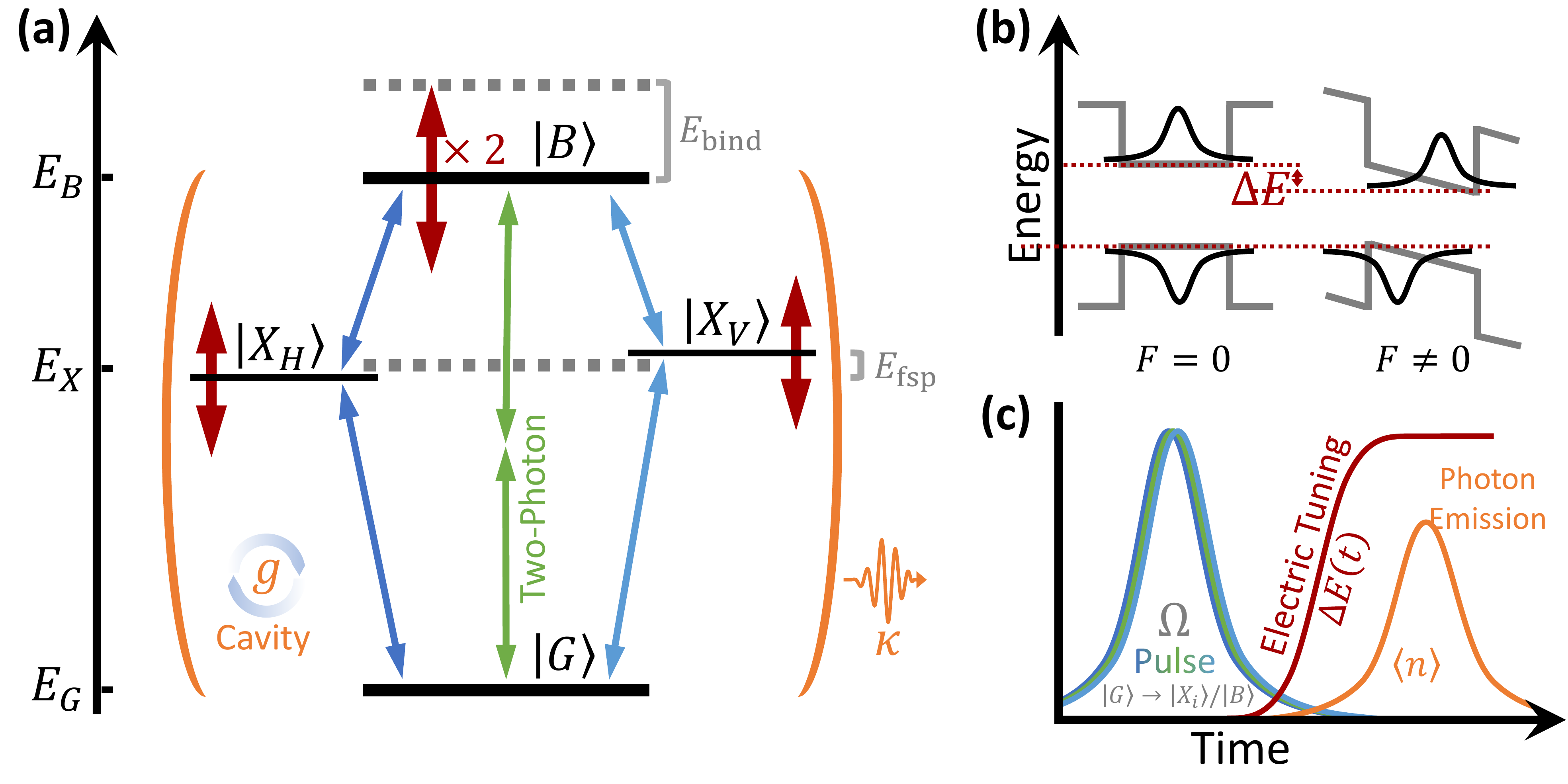}
\caption{\label{fig:physics}(a) QD-cavity system with electronic ground state $|G\rangle$, exciton states $|X_H\rangle$ and $|X_V\rangle$, and biexciton state $|B\rangle$ with binding energy $E_\text{bind}$ and single- and two-photon optical selection rules as indicated. The exciton state degeneracy may be lifted by fine structure splitting $E_\text{fsp}$. Electric tuning of exciton and biexciton energies via the quantum-confined Stark effect is also indicated. The shift in energy for the biexciton is assumed twice as large as for the excitons. (b) Illustration of the quantum-confined Stark effect that is induced for finite external electric field $F\neq0$. A time-dependent external field allows ultrafast control of the exciton and biexciton energies. (c) Sketch of the temporal sequence used for the coherent excitation pulse, electric tuning of electronic resonance frequencies, and resulting photon emission.}
\end{figure}
The proposed scheme does not use phonon side bands \cite{glassl2013proposed,quilter2015phonon}, optical Stark-shifts \cite{cosacchi2020demandOpticalStarkExcitation}, or the biexciton-exciton emission cascade.
Instead, ultrafast electric control \cite{WarburtonLodahl2020,PhysRevLett.84.733,warburton2000optical} of the electronic resonances based on the quantum-confined Stark effect is used. 
Coherent excitation of the quantum dot is done at finite dot-cavity detunings. The picosecond electric control of the exciton energies is then used to shift the exciton or two-photon biexciton resonance, respectively, towards zero dot-cavity detuning, the ideal condition for efficient photon emission.
In the present work we theoretically explore the potential and efficiency of this scheme with optoelectronic control, including the coupling of the dot-cavity system to its environment. 
The numerical results show that high single-photon and twin-photon emission probabilities are achieved with high values of single-photon indistinguishability and two-photon polarization entanglement, reaching near-unity values for ideal conditions where losses are minimized.

\section{\label{sec:model}Quantum Dot Model}

The theoretical description of the quantum-dot-cavity system starts with the four electronic configurations necessary to describe the optical excitations and emission dynamics discussed above. As sketched in \cref{fig:physics}(a), this includes the electronic ground state $\ket{G}$, two orthogonally polarized excitons $\ket{X_H}$ and $\ket{X_V}$, and the biexciton state $\ket{B}$. 
Coupling to a coherent classical light field and to two orthogonally polarized cavity modes with coupling strength $g$ is included. 
Interaction of system and environment includes photon losses from the cavity modes, radiative losses into non-cavity modes, pure dephasing, and coupling to acoustic phonons. 
Photon losses from the cavity modes occur with rate $\kappa$. 
The radiative loss of the dot population with $\gamma_\text{rad}=\ev{B}(T)\cdot\SI{1}{\mueV}$ varies with temperature through the averaged phonon displacement operator $\ev{B}$ (compare \cref{eqn:phoononbathdisplacement}). 
The pure dephasing rate of electronic states is given by $\gamma_\text{pure}=\SI{1}{\mueV/K\cdot T}$ and coupling to longitudinal acoustic phonons is included within a Lindblad-type contribution after applying a polaron transformation for finite and small temperatures $T\geq 0$ \cite{PhysRevB.85.115309,heinze2017polarization,PhysRevB.93.115308}. Full details on the theoretical approach are given in \cref{appendix:theory}.  \\
To allow for picosecond electric control, we model the effect of the quantum-confined Stark effect by varying the electronic energies in time as illustrated in \cref{fig:physics}(b) and (c). 
With the initial exciton energy $E_X$ and biexciton energy $E_B=2E_X-E_\text{bind}$, the time-dependent exciton and biexciton energies are given by $E_X \rightarrow E_X(t) = E_X(0) + \Delta E(t)$ and $E_B \rightarrow E_B(t) = E_B(0) + 2\Delta E(t)$. 
We note that the additional time dependency resulting from the electronic control $\Delta E(t)$ has to be included when calculating the polaron operators (compare \cref{eqn:polaronchi}). 
The small changes in oscillator strength that accompany the shifts in energy levels \cite{PhysRevLett.99.197403} and that can in principle be minimized \cite{ReimerPRL2019} are neglected here.
An exciton energy of $E_X=\SI{1.366}{eV}$ and biexciton binding energy of $E_\text{bind}=\SI{1}{\meV}$ are used and a small exciton fine structure splitting is included with $E_\text{fsp}=\SI{2}{\mueV}$ \cite{Heinze.2015,heinze2017polarization}. 
For all results shown in the present work, the length of optical pulses used for excitation is fixed to $\sigma=\SI{5}{ps}$, while both pulse area and frequency may vary as noted. \\

\begin{figure*}[t!]
\includegraphics[width=1.8\columnwidth]{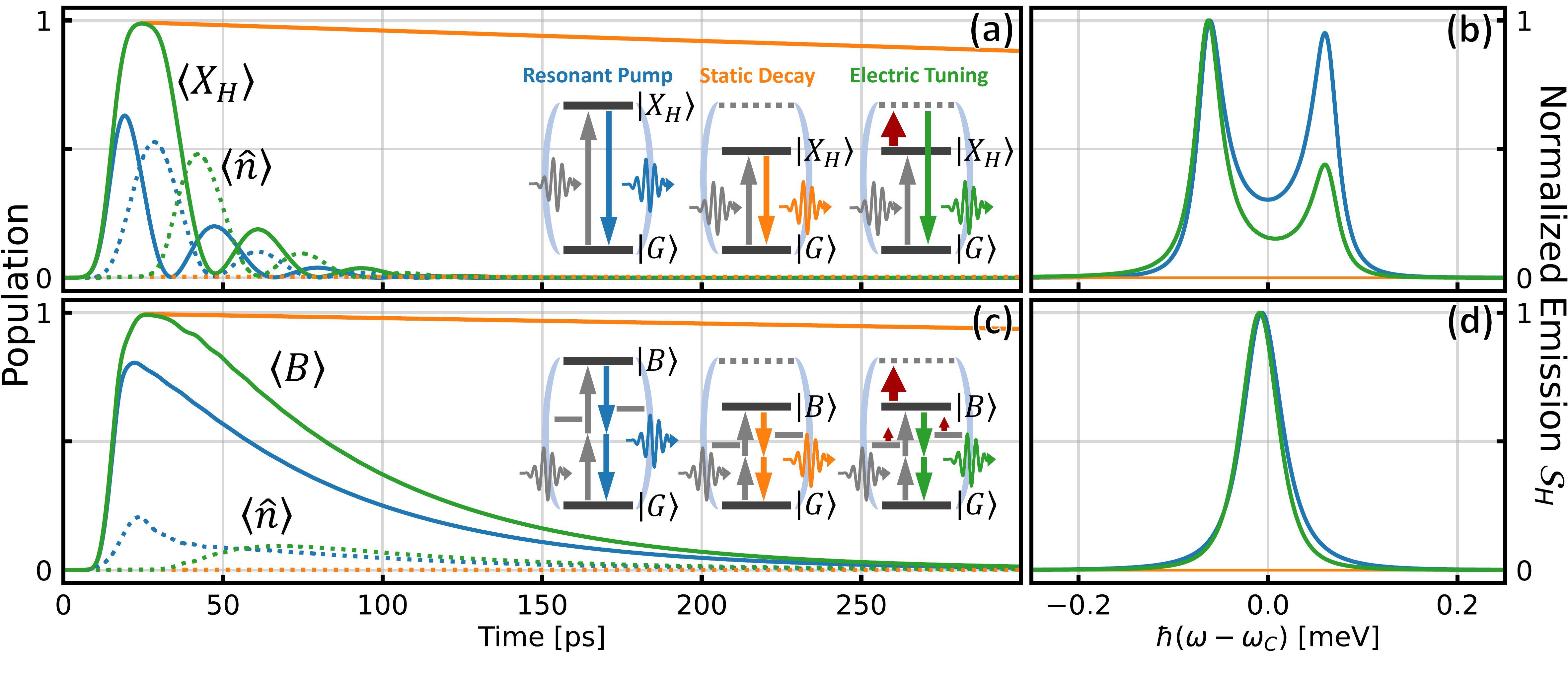}
\caption{\label{fig:basicdemo}Pulsed optical excitation of the quantum-dot-cavity system with and without ultrafast electric control. Two scenarios are shown, exciton excitation (top) and biexciton excitation (bottom) with exciton, biexciton (a,c, solid lines), and cavity photon populations given (a,c, dashed lines). The cavity loss is $\kappa=g$ with $g=\SI{66}{\mueV}$, resulting in a Q-factor of $\approx 20.000$. The excitation pulses are centered at $\SI{15}{ps}$ with a pulse width of $\sigma=\SI{5}{ps}$. The insets illustrate the three different configurations studied, the dashed lines mark the cavity resonance and the red arrows indicate the control of the electronic levels. In the electric tuning case a linear electronic control starting at $t_0=\SI{25}{ps}$ with total magnitude of $\Delta E = \SI{1}{\meV}$ transforms the static off-resonant case (orange) into the resonant case (blue) over a period of \SI{10}{ps}. Thus, in that case, excitation occurs while the electronic transitions are detuned from the cavity resonance and efficient photon emission starts at about $t_1=\SI{35}{ps}$ when the electric control restores the resonance condition with the cavity mode. This dynamic scenario (green) combines robust resonant exciton and biexciton initialization with the cavity-enhanced photon emission, which without the electronic control would be mutually exclusive. The resulting spectra in (b) and (d) inherit the slightly redshifted and asymmetric emission characteristics of the resonantly pumped case.} 
\end{figure*}

\section{\label{sec:results}Results}
In this section we will discuss the emission schemes used in the present work in more detail (\cref{sec:results:emissionscheme}) and analyze the numerical results obtained for the quantum-dot-cavity system introduced in the previous section for different cavity qualities and at different temperatures (\cref{sec:results:cavtempdep}).
\subsection{\label{sec:results:emissionscheme}Emission schemes}
First we would like to illustrate the general idea proposed. For this initial demonstration, only cavity losses are taken into account, no radiative decay, pure dephasing, or electron-phonon coupling are included.
As sketched in \cref{fig:basicdemo}, to lay the foundation for the later discussion, we investigate three different scenarios: (i) resonant optical pumping of the quantum dot exciton or biexciton, respectively, with also resonant emission into the cavity. Note that in a real system, direct pumping of photons into the cavity mode would occur in this case which for better comparison is not included here. (ii) Photon emission by spontaneous decay of an initially populated exciton or biexciton into an off-resonant cavity. In scenario (iii) the electronic transition energies are not fixed and electronic control is used to transition between cases (i) and (ii). Results for the three cases are summarized in \cref{fig:basicdemo}.

Starting from the system ground state, in case (i) the quantum dot is excited using the classical optical excitation pulse, such that the pulse frequency $\omega_L$ matches either the exciton resonance $\omega_L = \omega_X$ or the biexciton two-photon resonance $\omega_L = \omega_\text{2phot}$. 
For the excitation of a single exciton, a pulse area of $\Omega_0^X=\SI{1}{\pi}$ is used. For the biexciton, $\Omega_0^\text{2phot}=\SI{3.3}{\pi}$ is used. In this case the dot-cavity detuning is zero at all times, cf. blue sketch and lines in \cref{fig:basicdemo}. 
We note that in this case efficient cavity-resonant photon emission already starts occurring while the laser pulse keeps re-populating the excited states, the total emission probabilities (single-photon probability for exciton emission and two-photon probability for biexciton emission, respectively) surpasses $100\%$, reaching values of about $\mathcal{P}_\text{blue}\approx 110\%$ although a fully inverted electronic system is never reached.
To evaluate the quality of the photons emitted, besides the total emission probability for a given photon mode, defined in \cref{eqn:emissionprob} below, the following properties are analyzed: 
If the photon is the result of an exciton emission, the corresponding photon indistinguishability is calculated. 
For the twin-photon emission from the biexciton, the polarization entanglement is evaluated. 
These properties are considered ideal when reaching near-unity values, where higher is generally considered better. 
In the case (i) discussed, due to the ongoing excitation process interfering with the simultaneous decay of exciton or biexciton population, any photons emitted show significantly lowered values for the indistinguishability and polarization entanglement, respectively. 
In this case, the single-photon indistinguishability based on \cref{eqn:homindist} results in $\mathcal{I}_\text{blue} \approx 0.84$, and the twin-photon polarization entanglement based on \cref{eqn:conc} results in $\mathcal{C}_\text{blue} \approx 0.74$. 
Less than ideal values are obtained here even though our model does not account for direct generation of cavity-photons by the excitation pulse nor coupling to the environment except for the cavity losses. 
\cref{fig:basicdemo}(b) shows that the photons emitted feature the finite Rabi-splitting expected for the exciton emission and the usual Lorentzian emission characteristics for the direct biexciton two-photon emission \cite{ota2011spontaneous}. \\
\cref{fig:basicdemo} (orange inset and lines) also shows the results for case (ii). Here, the electronic transitions remain off-resonant with respect to the cavity mode with $\hbar(\omega_C-\omega_X)=\si{1}{\meV}$. 
In that case, photon emission into the off-resonant cavity is prolonged over several nanoseconds. The resulting emission probability on similar timescales when compared to the resonantly pumped case discussed above is much lower than anticipated for a useful on-demand photon source \cite{varnava2008good}, here reaching merely $\mathcal{P}_H\approx 10\%$ in the time frame shown. 
As no coupling to the environment except for cavity losses is included in these calculation, both the indistinguishability as well as the polarization entanglement remain at near-unity values with $\mathcal{I}_\text{orange} \approx 1$ and $\mathcal{C}_\text{orange} \approx 1$. 
Even with perfect initial state preparation, however, in an experimental implementation these values cannot be reached as the different loss mechanisms discussed above play an important role especially on such long emission timescales. 
In this off-resonant emission case where dot-cavity coupling is weak, photons emitted from both exciton and biexciton exhibit the usual Lorentzian spectral line shape (not visible in the spectral range shown in Fig.~2) as expected for the effectively almost cavity-less photon emission. \\
In case (iii) shown in \cref{fig:basicdemo} (green inset and lines) with finite initial dot-cavity detuning of $\hbar(\omega_C - \omega_X) = \SI{1}{\meV}$, an optical pulse is used to generate a high exciton or biexciton population, respectively. 
Immediately thereafter, the dot-cavity detuning is reduced over time using the electronic control, shifting the exciton or biexciton energy until $\omega_C - \omega_X = 0$, or $\omega_C - \omega_\text{2phot} = 0$, respectively, as sketched in \cref{fig:physics}(c). 
Numerically, a monotonic cubic interpolation between $\Delta E(t_0)$ and $\Delta E(t_1)$ is used to achieve a smooth, non-instantaneous transition between the two energy configurations. 
The dot-cavity detuning is reduced to zero with an average electronic control speed for both scenarios of $\left.\frac{d}{dt}\Delta E(t)\right\rvert_\text{avg}=\SI{100}{\mueV/ps}$, a value achievable in realistic structures. 
Therefore, zero dot-cavity detuning is reached after $\SI{10}{ps}$ for the exciton and $\SI{15}{ps}$ for the biexciton.  
The results shown in \cref{fig:basicdemo} (green lines) illustrate that using this electronic control of the resonance conditions, we can combine efficient excitation and efficient resonant emission. 
For the ideal case with no additional losses enabled, the electronic system can be initially fully inverted for both the exciton and the biexciton and is then shifted by the electric control. 
The excitation process is then followed by efficient resonant emission of photons into the cavity mode, with high emission probabilities, reaching near unity values in this idealized scenario without losses. 
The temporal emission characteristics resembles the initially resonant configuration (blue). Similar timescales are achieved, reducing the duration of the emission process to a picosecond timescale.
The total emission probability for the single-photon emission is equal to one, provided the system is initially fully inverted. For the twin-photon emission from the biexciton, the total emission probability is lower depending on the amount of electronic population lost to competing emission channels via the biexciton-exciton cascade. 
Any biexciton population lost to the exciton states does not contribute to the twin-photon emission, and will thus lower the photon yield from this process \cite{ota2011spontaneous}. 
We note that this strongly depends on the biexciton binding energy, where larger binding energies result in longer emission times due to the increased lifetime of the biexciton, which consequently lowers the efficiency of the twin-photon emission process due to the increased duration of the emission process.
Numerical values for the single-photon indistinguishability as well as the twin-photon polarization entanglement reach near unity values here, with $\mathcal{I}_\text{green}\approx 0.998$ and $\mathcal{C}_\text{green}\approx 0.994$. 
The numerical value calculated for the entanglement is very close to the theoretical maximum determined by the fine structure splitting \cite{cygorek2018comparison}, which with the parameters used in the present work is $\mathcal{C}_\text{max}=\frac{E_\text{bind}-E_\text{fsp}}{E_\text{bind}+E_\text{fsp}} = 0.996$.
This demonstrates that the electric control does not appear to significantly lower the efficiency of either emission process, suggesting that the degree of environmental coupling will eventually determine the achievable photon quality. 
Discussing the emission spectra, with the cavity initially being spectrally above the (bi-)exciton resonance, a notable redshift occurs for the single-photon emission, where either maximum of the Rabi-split peaks reached first by the electronic control is favored. 
Slower control ($\left.\frac{d}{dt}\Delta E(t)\right\rvert_\text{avg}\leq\SI{100}{\mueV/ps}$) will result in a more pronounced emission asymmetry, resulting in an overall larger redshift. 
Faster control will instead restore the usual resonant emission configuration faster, with an emission spectrum resembling the usual Rabi-split emission characteristics more closely. 
For the twin-photon emission, a similar yet much smaller redshift is observed. 
When mirroring the energy configuration such that the cavity resonance initially lies beneath the exciton or biexciton resonances, a corresponding blueshift will occur instead. 
We note that, when including coupling to the environment, this latter case would be more strongly affected by coupling to phonons, and both the indistinguishability as well as the polarization entanglement would be lowered significantly (not shown). Thus, results shown in the present work will be limited to those energy configurations reducing electron-phonon interactions.
\begin{figure}
\includegraphics[width=\columnwidth]{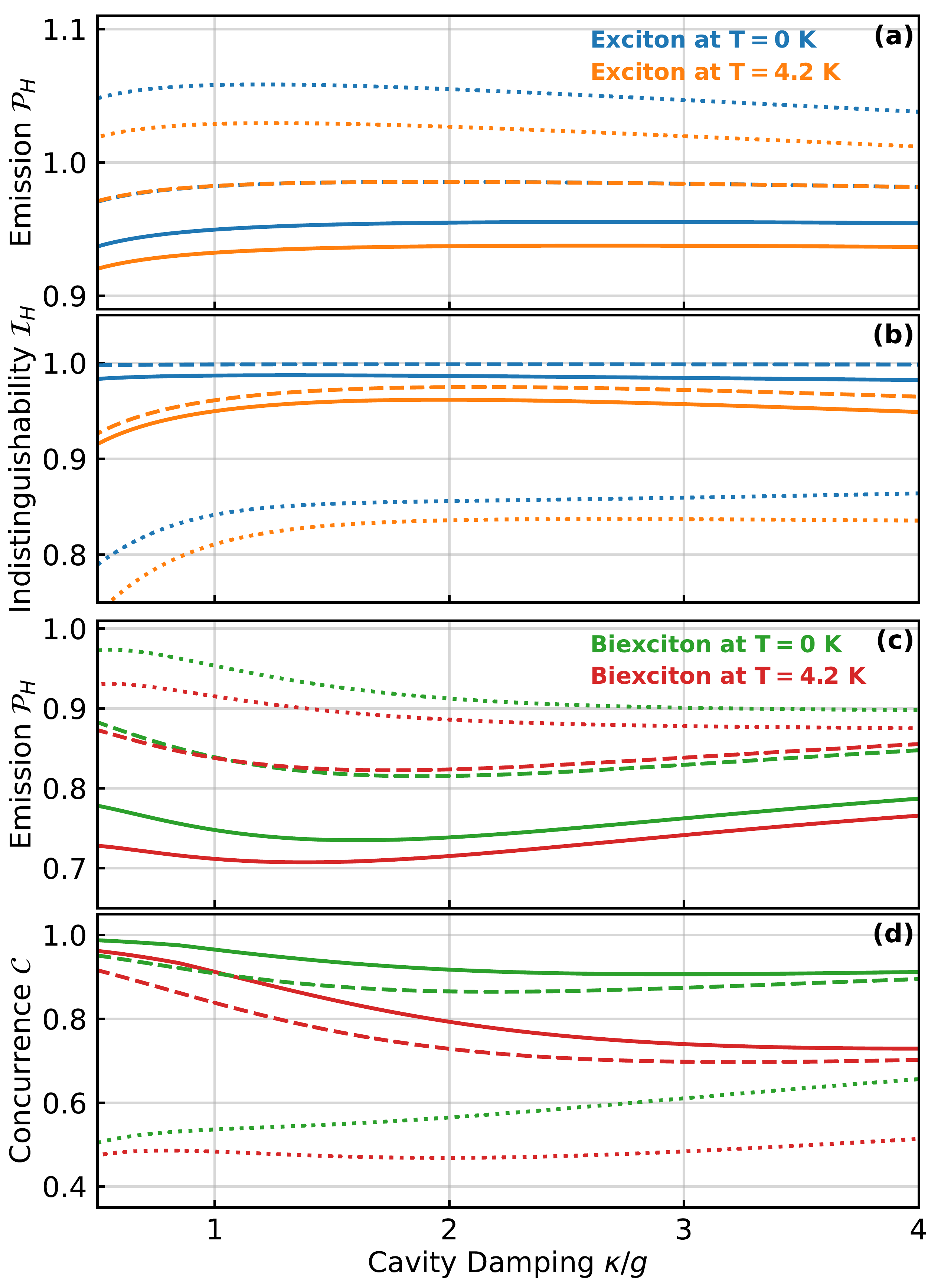}
\caption{\label{fig:dep_cavity}Emission probability for the horizontally polarized cavity photon (a,c) and the corresponding quantum properties (b,d) for different cavity loss $\kappa\in[0.5g,4g]$. For the photon emitted from the exciton, the single-photon indistinguishability is shown. For the photon emitted from the biexciton in a direct two-photon emission process, the concurrence as a measure for the polarization entanglement is displayed. The electrically controlled emission (solid lines) is compared with the spontaneous emission from the resonant decay of an initially fully excited exciton or biexciton, respectively, without the use of an excitation pulse (dashed lines) and resonantly pumped case as illustrated in \cref{fig:basicdemo} above (dotted lines).}
\end{figure}
\begin{figure}
\includegraphics[width=\columnwidth]{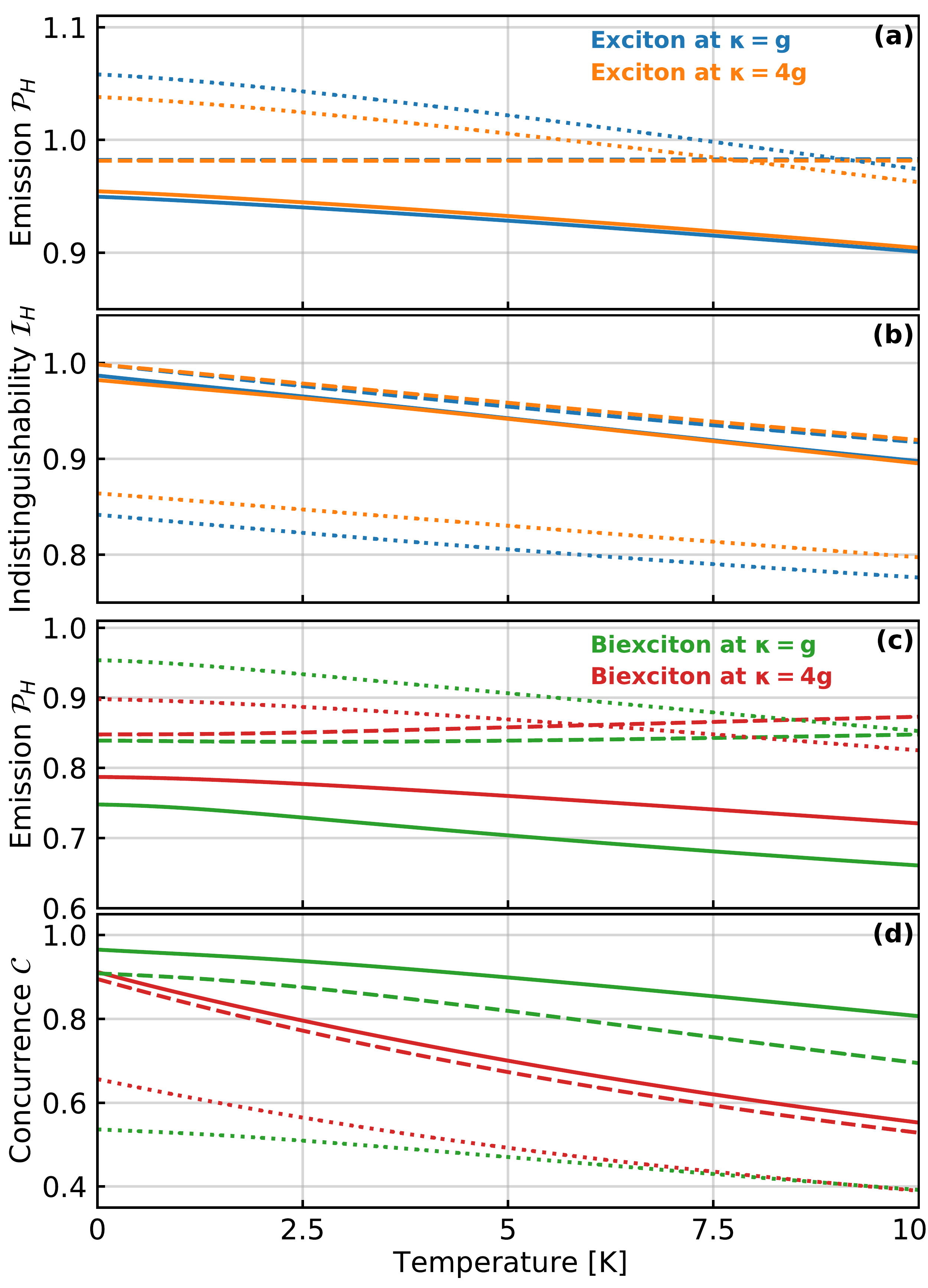}
\caption{\label{fig:dep_temperature}Emission probability for the horizontally polarized cavity photon (a,c) and the corresponding quantum properties (b,d) for different temperatures $T\in[\SI{0}{K},\SI{10}{K}]$. For the photon emitted from the exciton, the single-photon indistinguishability is shown. For the direct two-photon biexciton to ground state transition, the concurrence as a measure for the polarization entanglement is displayed. The electrically controlled emission (solid lines) is compared with the spontaneous emission from the resonant decay of an initially fully excited exciton or biexciton, respectively, without the use of an excitation pulse (dashed lines) and resonantly pumped case as illustrated in \cref{fig:basicdemo} above (dotted lines).}
\end{figure}
\subsection{\label{sec:results:cavtempdep}Cavity quality and temperature dependence}
\begin{figure}
\includegraphics[width=\columnwidth]{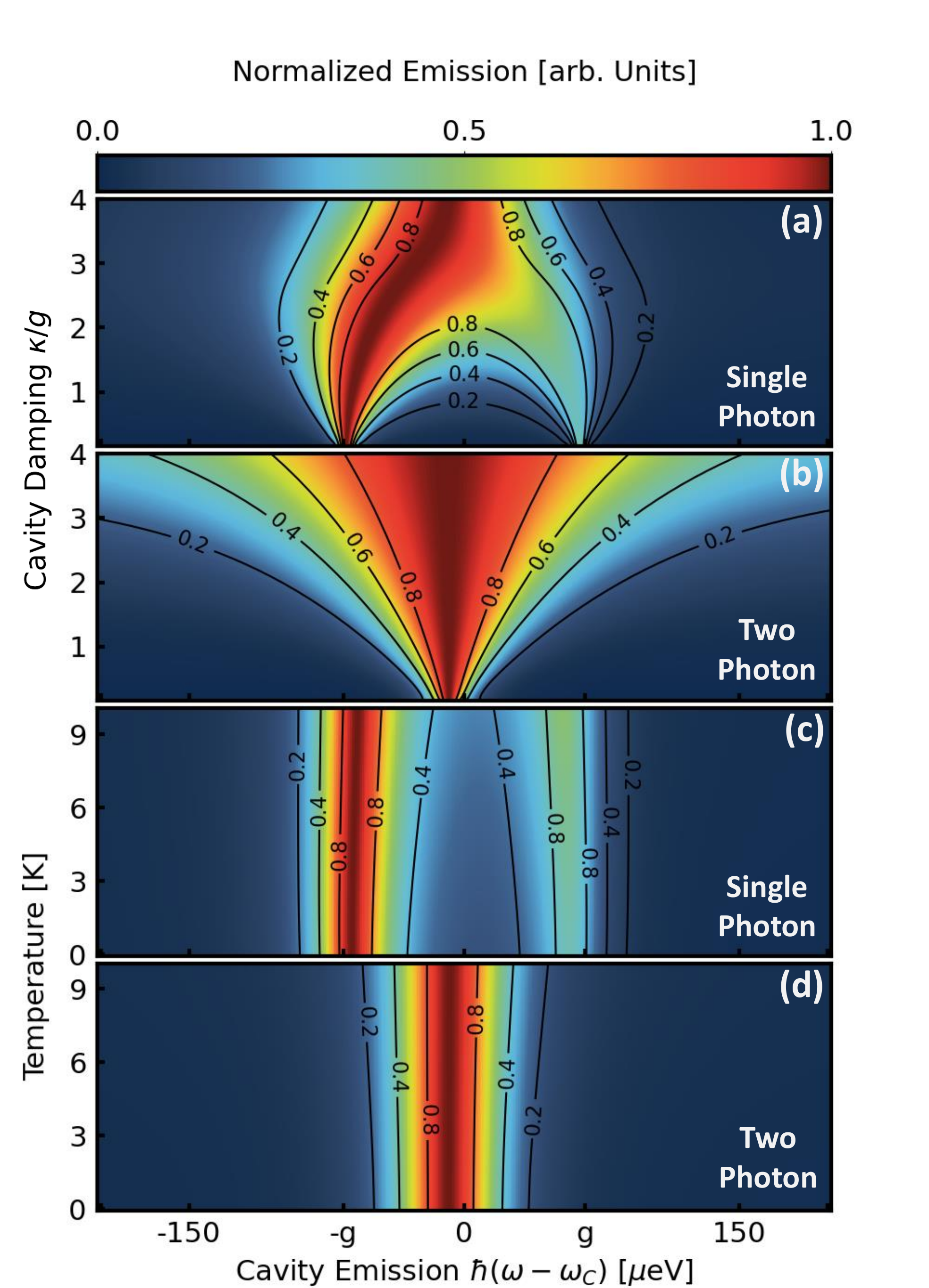}
\caption{\label{fig:plot_cavity_emission}Normalized emission spectra for the case with electric control for exciton (a,c) and biexciton (b,d). Parameters as in \cref{fig:dep_cavity}. For comparison (black lines), the emission spectra for the resonant decay case for initially fully excited exciton and biexciton, respectively, are overlayed. For the electrically controlled case, a redshifted photon emission from the exciton is observed, also showing an asymmetry for the Rabi-split emission peaks. Note that for mirrored energy configurations, the reverse control will result in a blue shift in emission energy instead.}
\end{figure}

In practical implementations, the cavity-enhanced emission from the exciton and the direct two-photon biexciton emission strongly depend on the quality of the cavity (Q-factor) used, as well as on the temperature of the semiconductor environment. 
In this section we include all loss mechanisms discussed above and present results for low temperatures and different cavity qualities using the configurations (i)-(iii) as discussed in \cref{sec:results:emissionscheme}.
The influence of different cavity loss rates $\kappa$ and different temperatures, on photon emission probability, single-photon indistinguishability, and the twin-photon polarization entanglement is investigated in detail. Results are summarized in \cref{fig:dep_cavity,fig:dep_temperature}.
%
%
The optical excitation pulse is fixed at either the exciton or at the two-photon resonance, respectively. 
For larger cavity losses, a small frequency shift for the resonance condition is expected for the optical transitions between the electronic levels \cite{PhysRevB.81.035302}. 
While this shift could in principle be included by careful design of the excitation pulse, this would sacrifice simplicity for only negligible improvement of the resulting photon emission probability, and is therefore not considered here. 
We note that the significantly lower excitation efficiency (compared to \cref{fig:basicdemo}) reached in \cref{fig:dep_cavity,fig:dep_temperature}, with also total emission probabilities for the static emission case (dashed lines) significantly higher than for the electronically tuned emission (solid lines), is not a result of this cavity-dependent shift in resonance condition. 
It is instead a consequence of the generally lower maximum excitation efficiency when including the environmental losses. Hence, the lower excitation efficiencies are not a consequence of the proposed emission scheme using the electronic control, but instead a generally encountered problem due to the environmental coupling.
While still remaining at relatively high values, the emission scheme with electronic control (\cref{fig:dep_cavity,fig:dep_temperature}, solid lines) exhibits slightly lower total single-photon and twin-photon emission probabilities of $\mathcal{P}_\text{X}\approx \SI{95}{\%}$ and $\mathcal{P}_\text{B}\approx \SI{75}{\%}$, respectively, compared to the ideal scenario with initially fully excited electronic states (dashed lines). 
This is due to the aforementioned imperfect excitation process when including additional environmental coupling as well as the additional time frame for the environmental coupling to influence the systems dynamics. 
In this context we further note that in the electric control scenario, smaller control speeds will result in a prolonged emission process, such that the system experiences more radiative decay, more pure dephasing and more dephasing due to phonon coupling. 
In that case the total photon emission probability as well as photon quality are generally reduced compared to the results shown in \cref{fig:dep_cavity,fig:dep_temperature}.\\
The emission probabilities for the twin-photon emission process show the expected behavior for varying either the degree of cavity losses or temperature, respectively. 
For the electrically controlled emission process, significant reductions in emission probability when compared to the emission from an initially fully excited biexciton can be seen in \cref{fig:dep_cavity,fig:dep_temperature}. 
This is the result of the losses in excitation efficiency for the static excitation pulse for the biexciton. The electric control by itself does not significantly reduce the emission probability.
The twin-photon polarization entanglement for the electrically controlled emission process surpasses the values achieved for the static, spontaneous emission from an initially fully excited biexciton state. This is a direct result of the initial conditions for the latter. 
No coherences $\rho_{G\leftrightarrow B}$ and $\rho_{B\leftrightarrow G}$ exist when numerically starting the time evolution in the biexciton state. The coherence generated by the optical pulse, however, is found to increase polarization entanglement. 
Analyzing the corresponding emission spectra displayed in \cref{fig:plot_cavity_emission} and comparing the static spontaneous emission to the electrically controlled emission, a redshift is observed for the latter scenario as already discussed for \cref{fig:basicdemo}. 
The faster the control the less pronounced this shift in emission energy, almost reproducing the characteristics of the static spontaneous emission in the limit of very fast electric control.

\section{\label{sec:conclusion}Conclusion}
We have presented the theoretical evaluation of a novel cavity-assisted emission scheme for single photons and photon pairs from a quantum dot exciton and biexciton state, respectively. 
This scheme combines the benefits of optical excitation off-resonant to the cavity mode with efficient cavity-resonant emission. 
The ultrafast transition between both is achieved employing the quantum-confined Stark effect that shifts the electronic resonances in and out off resonance with the cavity mode. 
We show that the limiting factor for achievable emission probability and quality of photons generated, remains the amount of coupling to the semiconductor environment. 
The faster the electronic control, however, the more its detrimental influence is reduced. 
Our numerical results include pure dephasing, photon losses, and electron-phonon coupling. These show that within the proposed scheme, single photons with high emission probability and near-unity indistinguishability as well as photon pairs with high emission probability and near ideal polarization entanglement can be generated in electrically controlled quantum dot-cavity systems. \\

\begin{acknowledgments}
This work was supported by the Deutsche Forschungsgemeinschaft (DFG) through the transregional collaborative research center TRR142 (grant No. 231447078, project A03) and Heisenberg program (grant No. 270619725) and by the Paderborn Center for Parallel Computing, PC$^2$. This project has received funding from the European Union's Horizon 2020 research and innovation program under grant agreement No. 820423 (S2QUIP) and No. 899814 (Qurope). We gratefully acknowledge fruitful discussions on the subject matter with Artur Zrenner. 
\end{acknowledgments}

\appendix*
\subsection{Theory} \label{appendix:theory}
To calculate the temporal evolution of the density matrix for the system visualized in \cref{fig:physics}, we numerically solve the von-Neumann equation,
\begin{align}
    \frac{\text{d}\rho}{\text{d}t} = \frac{\i}{\hbar}\left[\mathcal{H},\rho\right] + \sum \mathcal{L}_{\hat{O}}(\rho) \,, \label{eqn:vonneumann}
\end{align}
in matrix representation in Fock space. Here the last term includes the different contributions to the coupling of the QD-cavity system to its environment as detailed below. The Hamiltonian in the interaction picture and in the rotating frame reads:
\begin{align}
    \mathcal{H} = H^\text{I,RWA}_\text{QD-Cavity}+H^\text{I,RWA}_\text{QD-Light}~,
\end{align}
with the dot-cavity and dot-lightfield interaction parts:
\begin{align}
    H^\text{I,RWA}_\text{QD-Cavity} &= \sum_{i=H,V}g\left[ \ketbra{G}{X_i}\hat{b}_i^\dagger + \ketbra{X_i}{B}\hat{b}_i^\dagger \right] + \text{H.c.}~, \\
    H^\text{I,RWA}_\text{QD-Light} &= \sum_{i=H,V}\left[ \ketbra{G}{X_i}\Omega_i(t) + \ketbra{X_i}{B}\Omega_i(t) \right] + \text{H.c.}~,
\end{align}
with electronic ground state $\ket{G}$, exciton states $\ket{X_i}$, biexciton state $\ket{B}$, and cavity photon operators $\hat{b}_i^{\dagger}$.

For any operator $\hat{O}$, the interaction picture operator $\hat{O}^I$ is calculated by
\begin{align}
    \hat{O}^{\text{I}} = e^{i/\hbar \int_0^t H_0(t)\text{d}t}\hat{O}e^{-i/\hbar \int_0^t H_0(t)\text{d}t}~,
\end{align}
with 
\begin{align}
    H_0 =\sum_{\mathclap{i=G,X_H,X_V,B}}\;E_i\ketbra{i}{i} + \sum_{\mathclap{i=H,V}}\; E_c\hat{b}_i^\dagger\hat{b}_i~,
\end{align}
where $E_i$ are the (time dependent) energies of the electronic system defined in the main text. $E_c$ denotes the cavity mode energy for both optical modes. Since all operators are treated in the interaction frame, for readability, the superscript \textit{I} will be omitted in the following.

The external classical lightfield $\Omega_i(t)$ is defined as 
\begin{align}
    \Omega_i(t) = \frac{\hbar\Omega_0}{\sqrt{2\pi}\tau_0} \exp{-\frac{(t-t_0)^2}{2\tau_0^2} - \i\omega_i(t-t_0)}~,
\end{align}
with pulse area $\Omega_0$, frequency $\omega_i$, temporal width $\tau_0$ and time $t_0$. The coupling of the exciton-biexciton system to its semiconductor environment is included using a polaron transformation of the complete Hamiltonian of both parts. The resulting model then includes the interactions of the electronic states with a bath of longitudinal acoustic phonons. After the analytical treatment of the polaron transformed operators, a Lindblad-type contribution to the von-Neumann equation is obtained \cite{PhysRevB.85.115309}, with
\begin{widetext}
\begin{align}
    \mathcal{L}_{\text{Phonons}}\left[\rho(t)\right] = -\frac{1}{\hbar^2}\int_0^\infty \sum_{\mathclap{k=g,u}}\left(     G_i(\tau)\left[X_k(t),\tilde{X}_{k}(t,t-\tau)\rho(t-\tau)\right] +\text{h.c.}\right)\text{d}\tau~,
\end{align}
\end{widetext}
where
\begin{align}
    X_u &= \i\left(\Chi - \text{H.c.} \right)~, \\
    X_g &= \Chi + \text{H.c.}~.
\end{align}
The QD-bath interaction is then calculated by evaluating 
\begin{align}
    \Chi = \sum_{\mathclap{i=H,V}}\; \Big(\ketbra{X_i}{G} + \ketbra{B}{X_i}\Big)\left(g\hat{b}_i^\dagger + \Omega_i(t)\right). \label{eqn:polaronchi}
\end{align}
The polaron operator $\tilde{X}_i(t,t-\tau)$ is calculated by solving the von-Neumann equation \cref{eqn:vonneumann} with a reversed sign for the intitial condition $\tilde{X}_i(t,\tau=0)=\tilde{X}_i(t,t)$ until $\tilde{X}_i(t,t-\tau)$ is reached. 
Instead of the Lindblad terms, the explicit time dependency of $\tilde{X}_i(t,t-\tau)$ including the additional time dependency induced by the electric control of the dot energies is added onto the right side of the equation.

The Polaron-Green functions are
\begin{align}
G_g(\tau) &= \ev{B}^2\left(\cosh{(\phi(\tau))-1}\right)~, \\
G_u(\tau) &= \ev{B}^2\sinh(\phi(\tau))~,
\end{align}
with the averaged phonon displacement operator
\begin{align}
    \ev{B} = \ev{B_\pm}=\text{exp}\left[ -\frac{1}{2} \int_{0}^{\infty}\frac{J(\omega)}{\omega^2}\coth(\frac{\hbar\omega}{2\kb T})\text{d}\omega \right]~. \label{eqn:phoononbathdisplacement}
\end{align}
The phonon correlation function is given by: 
\begin{align}
    \phi(\tau) = \int_0^\infty\frac{J(\omega)}{\omega^2}\left[\coth\left(\frac{\hbar\omega}{2\kb T}\right)\cos\left(\omega\tau\right)-\i\sin\left(\omega\tau\right)\right]\text{d}\omega~.
\end{align}
These operators introduce the temperature dependency for the phonon emission and absorption.

The dynamics of the phonon bath are defined by their spectral distribution:
\begin{align}
    J(\omega) = \sum_{\vec{q}} \lambda_{\vec{q}}\delta\left(\omega-\omega_{\vec{q}}\right) = \alpha_p\omega^3e^{-\frac{\omega^2}{2\omega^2_b}}~.
\end{align}
The phonon cutoff energy is $\hbar\omega_b=\SI{1}{\meV}$. The phonon coupling factor is $\alpha_p = \SI{0.03E-24}{s^2}$.
Cavity losses, radiative decay as well as the electronic dephasing of the quantum dot population is included by the Lindblad-type contributions
\begin{align}
    \mathcal{L}_{\hat{O}}(\rho) = 2\hat{O}\rho\hat{O}^\dagger - \hat{O}^\dagger\hat{O}\rho - \rho\hat{O}^\dagger\hat{O}~. \label{eqn:lindblad}
\end{align}
For the cavity losses we have $\hat{O} = \sqrt{\kappa/2}\hat{b}_i$ with $\hbar\kappa = \SI{66}{\mueV}$ unless otherwise noted. For the radiative loss of the electronic population of the quantum dot we have $\hat{O} = \sqrt{\gamma_\text{rad}/2}\ketbra{X_i}{B}$. Population lost to this process will radiate into a non-cavity mode with $\hbar\gamma_\text{rad} = \expval{B}\SI{1}{\mueV}$ \cite{PhysRevX.1.021009}. Note that while this rate is never changed directly for different results, it is indirectly scaled by temperature due to the factor $\expval{B}$.
For the phonon induced pure dephasing of the electronic states we have $\hat{O} =\sqrt{\gamma_\text{pure}/8}(\ketbra{i}{i}-\ketbra{j}{j})$, with $\hbar\gamma_\text{pure} = \SI{1}{\mueV/K}$ \cite{PhysRevB.85.115309}, assumed to be proportional to the temperature of the environment.
%
%
To calculate the cavity emission spectrum, HOM-indistinguishability or two-photon concurrence, the first and second order photon correlation functions are calculated using the quantum regression theorem \cite{glauber1963quantum}:
\begin{align}
    G_i^{(1)}(t,t') &=\Tr{\rho'(t')\hat{b}_i^\dagger(0)} \label{eqn:g1} \\ &\text{with} \, \rho'(0) = \hat{b}_i(0)\rho(t)~, \nonumber \\
    G_{i,j}^{(2)}(t,t') &= \Tr{\rho'(t')\hat{b}_i^\dagger(0)\hat{b}_j(0)} \label{eqn:g2} \\
    &\text{with} \, \rho'(0) = \hat{b}_j(0)\rho(t)\hat{b}_i^\dagger(0)~. \nonumber
\end{align}
The correlation functions in \cref{eqn:g1,eqn:g2} are evaluated numerically by evolving the von Neumann equation, \cref{eqn:vonneumann}, for the corresponding $\rho'(0)$ initial condition.

The single-photon HOM-indistinguishability \cite{PhysRevB.98.045309} reads
\begin{widetext}
\begin{align}
    \mathcal{I}_{i} = 1-p_{c,i} = 1-\frac{\int_0^{t_\text{max}}\int_0^{{t_\text{max}}-t} 2G_{\text{HOM},i}^{(2)}(t,t')dt'dt}{\int_0^{t_\text{max}}\int_0^{{t_\text{max}}-t}\left( 2G_{\text{pop},i}^{(2)}(t,t')- \abs{\ev{\hat{b}_i(t+t')} \ev{\hat{b}_i^\dagger(t)}}^2 \right)dt'dt}~, \label{eqn:homindist}
\end{align} 
\end{widetext}
with 
\begin{align}
    G_{\text{HOM},i}^{(2)}(t,t') &= \frac{1}{2}\left( G_{\text{pop},i}^{(2)}(t,t') \right.\nonumber\\&+ \left.G^{(2)}_{i,i}(t,t') - \abs{G_i^{(1)}(t,t')}^2 \right)~, \\
    G_{\text{pop},i}^{(2)} &= \ev{\hat{b}_i^\dagger\hat{b}_i}(t)\ev{\hat{b}_i^\dagger\hat{b}_i}(t+t')~.
\end{align}
The two-photon concurrence is used as a measure for the polarization entanglement of the emitted photons. It is given by \cite{wootters1998entanglement}: 
\begin{align}
    \mathcal{C} = \text{max}\left\{{0,\lambda_4-\lambda_3-\lambda_2-\lambda_1}\right\}~, \label{eqn:conc}
\end{align}
where $\lambda_i$ are the numerical eigenvalues of
\begin{align}
    R&=\sqrt{\sqrt{\rho_\text{2ph}}\tilde{\rho}\sqrt{\rho_\text{2ph}}}~,
    \end{align}
with 
\begin{align}
    \tilde{\rho} &= (\sigma_y \otimes \sigma_y)\rho_\text{2ph}^*(\sigma_y \otimes \sigma_y)~.
\end{align}
Here, $\sigma_y$ is the corresponding spin-flip matrix and $\rho_\text{2ph}$ is the two-photon density matrix with 
\begin{align}
    \rho^\text{2ph}_{i,j} = \int_0^{t_\text{max}}\int_0^{t_\text{max}-t}G_{i,j}^{(2)}(t,t')dt'dt~.
\end{align}

The emission spectrum for either one of the cavity modes assuming ideal detection is calculated by \cite{Mirza_2014}:
\begin{align}
    \mathcal{S}_i(t_\text{max},\omega) = \Re \int_{0}^{t_\text{max}}\int_0^{t_\text{max}-t}G_i^{(1)}(t,t')e^{-\i\omega t'}\text{d}t'\text{d}t~. \label{eqn:ewspectrum}
\end{align}

The cavity emission probability $\mathcal{P}_i(t)$ is calculated by integrating the photon density times cavity loss rate: 
\begin{align}
\mathcal{P}_i(t) = \kappa\int_0^{t_\text{max}} \ev{\hat{b}_i^\dagger\hat{b}_i}(t)\text{d}t. \label{eqn:emissionprob}
\end{align}

\bibliography{bibo}

\end{document}